\documentstyle[aps,prb,twocolumn]{revtex}
\begin{document}
\title{Structure, elastic moduli and thermodynamics of sodium and potassium at
ultra-high pressures}
\date{Today}
\author{M. I. Katsnelson$^1$, G. V. Sinko$^2$, N. A. Smirnov$^2$, A. V.
Trefilov$^3$%
, and K. Yu. Khromov$^3$}
\address{$^1$Institute of Metal Physics, Ekaterinburg 620219, Russia\\
$^2$Federal Nuclear Center ``Institute of Technical Physics'', Snezhinsk\\
456770, Russia\\
$^3$Russian Science Center ''Kurchatov Institute'', Moscow 123182, Russia}
\maketitle
\draft

\begin{abstract}
The equations of state at room temperature as well as the energies of
crystal structures up to pressures exceeding 100 GPa are calculated for Na
and K . It is shown that the allowance for generalized gradient corrections
(GGA) in the density functional method provides a precision description of
the equation of state for Na, which can be used for the calibration of
pressure scale. It is established that the close-packed structures and BCC
structure are not energetically advantageous at high enough compressions.
Sharply non-monotonous pressure dependences of elastic moduli for Na and K
are predicted and melting temperatures at high pressures are estimated from
various melting criteria. The phase diagram of K is calculated and found to
be in good agreement with experiment.
\end{abstract}


The theoretical and experimental studies of the matter properties at
ultra-high pressures arouse a great interest in the connection with the
possibility to obtain phases with uncommon properties as well as geophysical
and astrophysical applications. As an example, the problem of metallic
hydrogen can be mentioned \cite{b1}. In the high pressure studies the alkali
metals can be conveniently used as model objects. This is due, first, to
their high compressibility and, second, to the variety of physical phenomena
occurring in their compression and numerous structural and electron phase
transitions (see, e.g.\cite{b2,b3,b4,b5,b5ad,b6,skriver,b7}). For heavy
alkali metals it is the famous $s-d$ isostructural FCC-FCC transition (see,
e.g., \cite{sd} and references therein) as well as the transitions to
uncommon distorted phases at higher pressures\cite{CsRb}. Recently it was
supposed, basing on the electron structure calculations, that lithium can
transform at high enough pressures into ``exotic'' phases similar to that of
hydrogen \cite{nature}. Thus, further theoretical investigations of
structural properties of alkali metals at ultra-high pressures seem to be
interesting and important.

Despite a lot of considerations, this is still an open problem. The most of
early attempts used computational approaches which were not accurate enough
from the contemporary point of view. It is well known (see, e.g., \cite{b8})
that the highly accurate quantitative description of the electronic and,
especially, lattice properties of metals needs the consideration of the real
form of potential in the crystal and going beyond the frame of local
approximation in the density functional, in particular, the allowance for
generalized gradient corrections (GGA) \cite{b9}. In the present work a
consistent theoretical study of the relative stability of crystal structures
of Na and K under pressure as well as a variety of related lattice
properties, is performed basing on these first-principle calculations. The
most interesting result obtained is that, contrary to the traditional
concepts (see, e.g., \cite{b7}) neither structure, which is characteristic
of metals under normal conditions (BCC, FCC and HCP), is stable at high
enough pressures even in Na where there are no electron transitions.

The {\it ab initio \/} calculations of electronic structure, thermodynamical
potential, equilibrium lattice parameters and elastic moduli at temperature $%
T=0$ were carried out using the FP-LMTO method \cite{b10} with allowance for
the GGA in the form proposed in \cite{b9}. A careful optimization of the
parameters of this method \cite{b10} made it possible to carry out the
calculations of the total energy with an accuracy within the limits of 0.1
mRy/atom. Parameter $c/a$ for the HCP lattice was determined by the
minimization of the total energy for a fixed specific volume, and the
elastic moduli --- by numerical differentiation of the total energy with
respect to tetragonal and trigonal deformations (see, e.g., \cite{b7}). Up
to now, there are only few works devoted to the first-principle calculations
of elastic moduli of metals under high pressures (see, e.g.,\cite{WMo} for
Mo and W). The expressions for the free energy $F$ (connected with Gibbs
thermodynamical potential $G$ by the relationship $G=F+PV$, $P=-dF/dV$,
where $P$ is the pressure, $V$ is the volume) and elastic moduli $C_{{\alpha
\beta \gamma \delta }}$ at finite temperatures can be presented in the
following form:
\begin{equation}
F(V,T)=E_{{e\ell }}(V)+F_{{ph}}(V,T)  \label{eq1}
\end{equation}
\begin{eqnarray}
C_{{\alpha \beta \gamma \delta }}(V,T) &=&C_{{\alpha \beta \gamma \delta }%
}^{0}(V_{0})+V_{0}\frac{dC_{{\alpha \beta \gamma \delta }}^{0}}{dV_{0}}\frac{%
P_{{ph}}(V_{0},T)}{B_{0}}  \nonumber  \label{eq2} \\
&&\ +C_{{\alpha \beta \gamma \delta }}^{\ast }(T)  \label{eq2}
\end{eqnarray}
where $F_{{ph}}=T\sum\limits_{{\xi ,\vec{q}}}\ell n\left[ 2sh\frac{\hbar
\omega _{_{{\xi }}}(\vec{q})}{2T}\right] $ is the free energy of the phonon
subsystem in the harmonic approximation, $E_{{e\ell }}(V)$ is the total
energy of the electron subsystem obtained from the FP-LMTO calculations \cite
{b10}, $P_{{ph}}=-\partial F_{{ph}}/\partial V$ is the phonon pressure, $%
B_{0}$ is the bulk modulus of the electron subsystem at $T=0$, $\xi ,\vec{q}%
,\omega $ are the number of phonon branch, wave vector and phonon frequency,
respectively. In the expression for the elastic moduli $C_{{\alpha \beta
\gamma \delta }}(V,T)$ the first term corresponds to the electron
contribution at $T=0$, second one - to the quasiharmonic contribution due to
the effects of thermal expansion, and third one - to the phonon contribution
obtained from the differentiation of $F_{{ph}}$ with respect to the
corresponding deformation parameters. For the calculation of phonon
contributions to thermodynamical functions the pseudopotential model
described in \cite{b11}, which describes with a high accuracy a wide range
of lattice properties of alkali metals, was used.

Figs. 1,2 show the results of calculations of Gibbs potentials at $T=O$ for
the BCC, FCC and HCP phases of sodium and potassium, respectively. It should
be noted that in the case of Na the phonon contribution to $\triangle G$
(contribution of zero-point vibrations, $\triangle G_{{zp}}$) are $G_{{zp}%
}^{fcc}-G_{{zp}}^{bcc}=1.31\times 10^{-5}$ Ry/atom, $G_{{zp}}^{hcp}-G_{{zp}%
}^{bcc}=1.35\times 10^{-5}$ Ry/atom. This is well comparable with the
electron contributions to $\triangle G$ under the normal conditions.
However, already at $P>1$ GPa for Na and practically at all the pressures
for K the contribution of zero-point oscillations to $\triangle G$ can be
neglected. Generally speaking, energy differences of order of $10^{-5}$
Ry/atom is too small to be accurately derived in our first-principle
calculations; nevertheless, we have obtained correct phase diagram even for
sodium at low pressures. In accordance with the results of calculations, at $%
P=0$ the BCC phase and HCP phase have the lowest energy for K and Na,
respectively (actually, under these conditions Na has not HCP but 9R
structure whose energy, however, is very close to HCP \cite{b7}). It is
important to emphasize that this difference of Na from K is purely
quantitative: according to the results, shown in the insert to Fig.2, K
would have to transit to the hexagonal close-packed phase at the negative
pressure of the order of several kilobars. As the calculations show
potassium, unlike sodium, transits from the BCC to FCC structure at $%
P\approx 11.6$ GPa, which is in excellent agreement with the experimental
data \cite{b5,b5ad}. In this case the relative change in the volume $%
\triangle =(V_{{bcc}}-V_{{fcc}})/V_{{bcc}}\approx 0.0067$ takes place at $%
V_{0}/V=2.14$. Here and below $V_{0}$ is the experimental value of the
specific volume at the atmospheric pressure and temperature 10K, equal to
484.12 a.u. \cite{b12}. Our calculations make it possible to suppose that
the difference between Na and K is associated with the electronic
topological transition occurring in the BCC potassium at $V_{0}/V\approx 2$
and destabilizing the BCC structure. The similar situation takes place in Li
\cite{b7} while in the BCC sodium, within the whole range of pressures, the
Van-Hove singularity goes away from the Fermi level under the compression.
Generally, sodium seems to be a unique metal in the Periodical Table: in the
whole region of the existence of BCC structure it has no singularities of
electronic structure near the Fermi level, and the Fermi surface remains
approximately spherical.

The calculation results of equations of state for sodium and potassium are
shown in Fig.3 along with the experimental data available. It should be
pointed out that the experimental data agree with the theoretical results
within their accuracy limits ($\approx 10\%$). This creates the
prerequisites for the development of pressure scale based on sodium as a
reference substance. Note also that at room temperature the role of the
phonon contribution to pressure falls under the compression, and this
contribution in itself is small (of the order of 0.3 GPa at full pressure
20-30 GPa).

Figs. 4,5 display the calculation results of the dependence of elastic
moduli $C_{{ij}}(V)$ on the compression for sodium and potassium,
respectively. A drastically non-monotonous behavior of shear moduli
associated with the tetragonal ($C^{\prime }=(C_{11}-C_{12})/2$) and
trigonal ($C_{44}$) deformations in both BCC and FCC structures is
noticeable. It should be pointed out that at least at compressions $%
V_{0}/V<2 $ the calculation results of $C_{{ik}}(V)$ in the pseudopotential
model and in the first-principle approach are close. In this region the
equations of state coincide in these two approaches with an accuracy up to
several percents. This confirms a sufficiently high reliability of our use
of pseudopotential model for the calculation of phonon contributions to
thermodynamical values. Nevertheless, the phonon contributions to the shear
moduli do not exceed 10\% within the whole pressure range studied. It should
be noted that softening of modulus $C^{\prime }$ is a typical pre-transition
phenomenon connected with the structural transitions between the BCC and
close-packed structures\cite{b13}. However, the softening of modulus $C_{44}$
in the FCC structure of K at high pressures (Fig.5) is rather surprising. It
appears to be similar to the softening of this modulus, taking place in the
FCC structure of Cs near the electron $s\rightarrow d$ transition\cite{b14}
and is due to the crawling of the Fermi level over the peak of $d$-state
density.

Fig.6 shows an experimental phase diagram of potassium and the phase diagram
built on the basis of our calculations. The dependence of the melting
temperature on pressure, $T_m(P)$ in the BCC and FCC phases was obtained
using the phonon spectra and different melting criteria. First of all, we
use the Lindeman criterion
\begin{equation}
\overline{x^2(T_m)}/d^2=const,  \label{eq3}
\end{equation}
where $\overline{x^2(T)}=\sum\limits_{{\xi \vec q}}\frac{\hbar \left| \vec q%
\vec e_{{\xi \vec q}}\right| ^2}{2M\omega _{{\xi \vec q}}}\coth \frac{\hbar
\omega _{{\xi \vec q}}}{2T}$ is the mean square of atom displacement, $\vec e%
_{{\xi \vec q}}$ is the polarization vector, $M$ is the atom mass, $d$ is
the distance between the nearest neighbors. Although the Lindeman criterion
is empirical, it may be expected that its use for finding the melting
temperature at high pressures would be as successful as at low temperatures
\cite{b15}. Nevertheless one can see from Fig.6 that it is not too accurate
in a broad pressure region. Varshni melting criterion \cite{v} which is
based on the temperature softening of the shear moduli, namely
\begin{equation}
C_{44}\left( T_m\right) /C_{44}\left( 0\right) =0.65,  \label{eq4}
\end{equation}
appears to be much more accurate. Here we use the method of the calculation
of the temperature dependence of elastic moduli from the phonon spectra
described in \cite{b3}. Note also that we describe with high accuracy the
BCC-FCC phase boundary. We also present the results obtained in generalized
Debye model \cite{sk} when all the thermodynamical quantities are calculated
in the Debye model but with the Debye temperature found from {\it ab initio}
elastic moduli. One can see that this description is also rather accurate
for potassium.

Fig.1 shows that the BCC phase of sodium becomes energetically unfavorable
as compared with the HCP at a pressure about 80 GPa. At $P>100$ GPa,
however, this phase demonstrates anomalies in the equilibrium value of
parameter $c/a$ (Fig.7). A sharp decrease in the ratio $c/a$ to 1.2-1.3 at $%
V_{0}/V>4.35$, which is necessary to maintain the HCP lattice in
equilibrium, is doubtful actually, and seems to be indicative of transition
to some non closed packed phase with a large number of atoms per cell. These
phases are observed in K, Rb and Cs at high pressures \cite{b5,b5ad,CsRb}.
It is usual to associate their appearance in heavy alkali metals with the $%
s\rightarrow d$ transition. Thus, according to our results, all the three
''typically metallic'' structures - BCC, FCC, HCP do not correspond to the
lowest energy in Na where no electronic transitions are observed within the
pressure range considered. In order to understand qualitatively the cause of
appearance of ''nonstandard'' metallic phases, let us use the above
mentioned pseudopotential model for the estimations. In this model the
radius of ''hard'' ion core is described by the pseudopotential parameter $%
r_{0}$ \cite{b11}. As the estimates show the compression $V_{0}/V\approx 4$
corresponding to the instability of the close- packed phase coincides with
the condition of overlapping ion cores $2r_{0}\approx d$ for Na. Hence, the
concept of well determined ion cores, being at the base of standard metallic
bond description, becomes inapplicable at ultra-high pressures. As a result,
as we have seen, the substance transforms into exotic non close-packed
phases. These results are in qualitative agreement with the results\cite
{nature} for lithium.

In conclusion, note that it would be interesting to study the structure of
sodium at ultra-high pressures, which, as follows from the results obtained,
may prove to be surprising. Another result of this work, permitting a direct
experimental check is non-monotonous behavior of Na and K shear moduli at
pressure. At last, precision theoretical description of the equation of
state of sodium would make possible to use it for the development of an
accurate pressure scale up to 100 GPa. Although the contemporary
first-principle calculations can provide high enough accuracy also for
another substances (see, e.g., recent calculations\cite{Si} for Si) a very
high compressibility of sodium and the absence of phase transitions in a
broad range of pressures make it probably the most suitable for these
purposes.

The authors are grateful to D. Yu. Savrasov and S. Yu. Savrasov for the
permission to use the author's version of the code realizing the method \cite
{b10} in their work as well as to D. Yu. Savrasov and E. G. Maksimov for
useful discussions of the details of this method.

\newpage

Figure captions

Fig.1. Pressure dependence of the differences of Gibbs potentials between
BCC and FCC as well as HCP and FCC structures for Na.

\bigskip Fig.2. Pressure dependence of the differences of Gibbs potentials
between BCC and FCC as well as HCP and FCC structures for K.

\bigskip Fig.3. Equations of states for sodium and potassium at $T=295K$.
Solid line corresponds to FP-LMTO calculations \cite{b10}, dashed line - to
the calculations by the pseudopotential method \cite{b11}, Empty (solid)
triangles, circles and asterisks (squares) are the experimental data \cite
{b16,b17,b18,b5ad} for Na and K, respectively.

\bigskip Fig.4. The dependence of elastic moduli $C^{\prime }$ and $C_{{44}}$
on the compression $U$ for BCC Na; empty circles and squares show,
respectively, the data from \cite{b2}; the solid ones - the values obtained
in the present work.

\bigskip Fig.5. The dependence of elastic moduli $C^{\prime }$ and $C_{{44}}$
on the compression $U$ for BCC and FCC phases of K. Solid (empty) circles
and squares show $C^{\prime }$ and $C_{{44}}$ values for BCC (FCC) phases,
respectively. The dashed line --- the $C^{\prime }$ values for the bcc phase
from \cite{b2}.

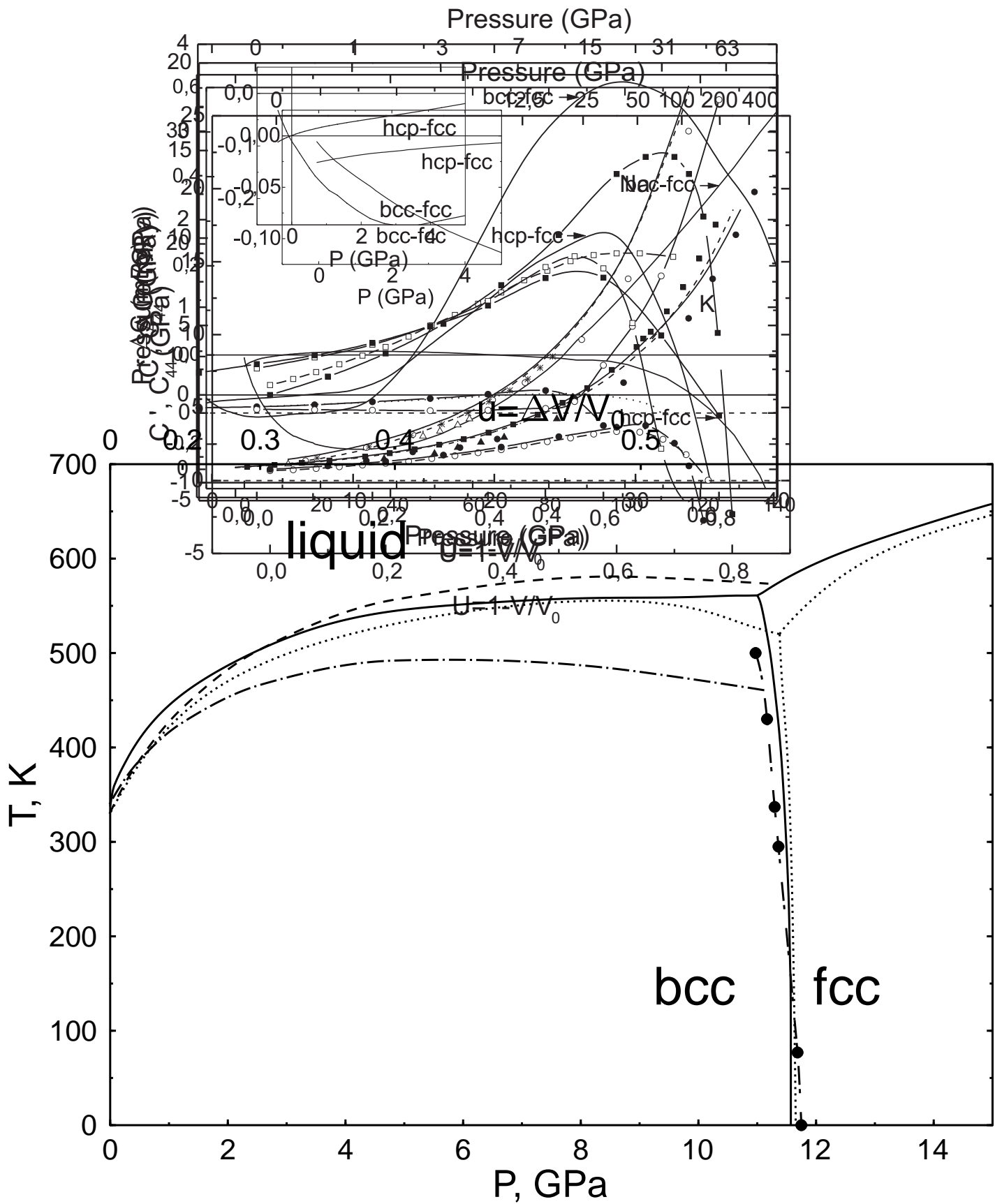
\bigskip Fig.6. Phase diagram of potassium. The solid line - experimental
data \cite{b6}, dashed line- the calculations using Varshni criterion (\ref
{eq4}) dashed-dot line - the calculations using Lindeman criterion (\ref{eq3}%
), dotted line - generalized Debye model (see the text). Solid circles -
BCC-FCC\ phase boundary from our calculations.

\bigskip Fig.7. Dependence of the total energy of HCP structure for Na on
the ratio $c/a$ for various compressions: the solid line --- $U=0.75$;
dashed line --- $U=0.76$; dashed-dot line --- $U=0.765$. The insert shows
the equilibirum values of the parameter $c/a$ for HCP structure depending on
$U$ is shown. Solid (empty) circles denote the values taken in the global
(local) minimum of the total energy, correspondingly.

\end{document}